\documentclass[preprint, 3p, twocolumn, letterpaper, 10pt]{elsarticle} 
\usepackage[hyphens]{url}

  \journal{Variance} 

\usepackage{lineno} 
\providecommand{\tightlist}{%
  \setlength{\itemsep}{0pt}\setlength{\parskip}{0pt}}

\usepackage{graphicx}
\usepackage{booktabs} 

\usepackage[T1]{fontenc}
\usepackage{lmodern}
\usepackage{amssymb,amsmath}
\usepackage{ifxetex,ifluatex}
\usepackage{fixltx2e} 
\IfFileExists{upquote.sty}{\usepackage{upquote}}{}
\ifnum 0\ifxetex 1\fi\ifluatex 1\fi=0 
  \usepackage[utf8]{inputenc}
\else 
  \usepackage{fontspec}
  \ifxetex
    \usepackage{xltxtra,xunicode}
  \fi
  \defaultfontfeatures{Mapping=tex-text,Scale=MatchLowercase}
  
\fi
\IfFileExists{microtype.sty}{\usepackage{microtype}}{}
\usepackage{longtable}
\ifxetex
  \usepackage[setpagesize=false, 
              unicode=false, 
              xetex]{hyperref}
\else
  \usepackage[unicode=true]{hyperref}
\fi
\hypersetup{breaklinks=true,
            bookmarks=true,
            pdfauthor={},
            pdftitle={Towards Explainability of Machine Learning Models in Insurance Pricing},
            colorlinks=false,
            urlcolor=blue,
            linkcolor=magenta,
            pdfborder={0 0 0}}
\usepackage{booktabs}
\usepackage{longtable}
\usepackage{array}
\usepackage{multirow}
\usepackage{wrapfig}
\usepackage{float}
\usepackage{colortbl}
\usepackage{pdflscape}
\usepackage{tabu}
\usepackage{threeparttable}
\usepackage{threeparttablex}
\usepackage[normalem]{ulem}
\usepackage{makecell}
\usepackage{xcolor}

\begin{document}
\begin{frontmatter}

  \title{Towards Explainability of Machine Learning Models in Insurance Pricing}
    \author[a]{Kevin Kuo\corref{c1}}
   \ead{kevin@kasa.ai} 
   \cortext[c1]{Corresponding Author}
    \author[b]{Daniel Lupton}
   \ead{dlupton@taylorandmulder.com} 
  
      \address[a]{Kasa AI, 3040 78th Ave SE \#1271, Mercer Island, WA 98040, USA}
    \address[b]{Taylor \& Mulder, 10508 Rivers Bend Lane, Potomac, MD 20854, USA}
  
  \begin{abstract}
  Machine learning methods have garnered increasing interest among actuaries in
  recent years. However, their adoption by practitioners has been limited, partly
  due to the lack of transparency of these methods, as compared to generalized
  linear models. In this paper, we discuss the need for model interpretability in
  property \& casualty insurance ratemaking, propose a framework for explaining
  models, and present a case study to illustrate the framework.
  \end{abstract}
  
 \end{frontmatter}

\hypertarget{introduction}{%
\section{Introduction}\label{introduction}}

Risk classification for property \& casualty (P\&C) insurance rating has traditionally
been done with one-way, or univariate, analysis techniques. In recent years, many
insurers have moved towards using generalized linear models (GLM), a multivariate
predictive modeling technique, which addresses many shortcomings of univariate
approaches, and is currently considered the gold standard in insurance risk
classification. At the same time, machine learning (ML) techniques such as deep
neural networks have gained popularity in many industries due to their superior
predictive performance over linear models (LeCun, Bengio, and Hinton 2015). In fact, there
is a fast growing body of literature on applying ML to P\&C reserving
(Kuo 2019; Wüthrich 2018; Gabrielli, Richman, and Wüthrich 2019; Gabrielli 2019). However, these ML techniques, often considered to be
completely ``black box'', have been less successful in gaining adoption in pricing,
which is a regulated discipline and requires a certain amount of transparency in models.

If insurers can gain more insight into how ML models behave in risk classification
contexts, it would increase their ability to reassure regulators and the public
that accepted ratemaking principles are met. Being able to charge more accurate
premiums would, in turn, make the risk transfer system more efficient and contribute
to the betterment of society. In this paper, we aim to take a step towards liberating
actuaries from the confines of linear models in pricing projects, by proposing a
framework for explaining ML models for ratemaking that regulators, practitioners,
and researchers in actuarial science can build upon.

The rest of this paper is organized as follows: Section \ref{ratemaking} provides
an overview of P\&C ratemaking, Section \ref{need} discusses the importance of
interpretation, Section \ref{interpretability} discusses model interpretability
in the context of ratemaking and proposes specific tasks for model explanation,
Section \ref{application} describes current model interpretation techniques and
applies them to the tasks defined in the previous section, and Section
\ref{conclusion} concludes.

\hypertarget{ratemaking}{%
\section{Property and Casualty Ratemaking}\label{ratemaking}}

\hypertarget{history-of-ratemaking}{%
\subsection{History of Ratemaking}\label{history-of-ratemaking}}

Early classification ratemaking procedures were typically univariate in nature.
For example, Lange (1966) notes that (at that time) most major lines of insurance
used univariate methods based around the same principle: distributing an overall
indication to territorial relativities or classification relativities based on
the extent to which they deviated from the average experience.

Bailey and Simon (1960) introduced minimum bias methods, which were expanded
throughout the 60s, 70s, and 80s. As computing power developed, minimum bias
began to give away to GLMs, with papers such as Brown (1988) and Mildenhall (1999)
bridging the gap between the methods.

Arguably, GLMs predate minimum bias procedures by a significant margin. The term
was coined by Nelder and Wedderburn (1972), but generalizations of least squares
linear regression date back at least to the 1930s. Like minimum bias methods,
GLMs did not become mainstream in actuarial science for some time. For example,
the syllabus of basic education of the Casualty Actuarial Society (CAS) does not
seem to include any mention of GLMs
prior to Brown (1988) in the 1990 syllabus. From there, GLMs
seem to have received only passing mention until 2006 with the introduction of
Anderson et al. (2005) to the syllabus. Beginning in 2016, the CAS introduced Goldburd, Khare, and Tevet (2016)
to the syllabus, which offers a comprehensive guide to GLMs.

\hypertarget{machine-learning-in-ratemaking}{%
\subsection{Machine Learning in Ratemaking}\label{machine-learning-in-ratemaking}}

Paralelling the development of GLM was the development of machine learning
algorithms throughout the middle part of the 20th century. Detailed histories
of machine learning may be found in sources such as Nilsson (2009) and
Wang and Raj (2017). Consistent with GLMs, machine learning was relatively unpopular
in actuarial science until the last ten years as computing power has become cheaper
and more easily available and as machine learning software packages have obviated
the need for developing analyses from scratch each time an analysis is performed.
Due to the breadth of machine learning as a field, it is difficult to identify
the first time it entered the CAS syllabus; however, cluster analysis (in the
form of k-means) seems to have been first included in 2011 with Robertson (2009).
More recently, the CAS MAS-I and MAS-II exams introduced in 2018 have included
machine learning explicitly.

Within the area of ratemaking, machine learning is still in its infancy. A
significant portion of machine learning applications to ratemaking has been in
the context of automobile telematics, such as Gao, Meng, and Wuthrich (2018), Gao and Wuthrich (2018), Gao and Wuthrich (2019),
Roel, Antonio, and Claeskens (2018), or Wuthrich (2017). Presumably this focus has been a result of the
high-dimensionality and complexity of telematics data, making it a field in
which the unique abilities of machine learning techniques give a clear
advantage over traditional approaches.

Outside of telematics, Yang, Qian, and Zou (2018) use a gradient tree-boosting approach to
capture non-linearities that would be a challenge for GLMs. Henckaerts et al. (2018)
make use of generalized additive models (GAM) to improve predictions of GLMs.
Many researchers, in an apparent effort to demonstrate the range of possibilities
and advantages of machine learning, have approached the topic by comparing many
different machine learning algorithms within a single study, such as in
Dugas et al. (2003), Noll, Salzmann, and Wuthrich (2018), Spedicato, Dutang, and Petrini (2018). These studies make use of such varied
techniques as regression trees, boosting machines, support vector machines, and
neural networks.

\hypertarget{ratemaking-process}{%
\subsection{Ratemaking Process}\label{ratemaking-process}}

Regardless of the method employed for determining this risk of various
classifications, the actual process of setting rate relativities typically
involves some variation of the following steps:

\begin{enumerate}
\def\labelenumi{\arabic{enumi}.}
\tightlist
\item
  Obtain relevant policy-level data
\item
  Prepare data for analysis
\item
  Perform analysis on the data, employing desired method(s) to estimate
  needed rates
\item
  Select final rates based on rate indications
\item
  Present rates to the reglator, including explanation of the steps followed to
  derive the rates
\item
  Answer questions from regulators regarding the method employed
\end{enumerate}

The focus of this paper is on steps 5 and 6.
In many states, rate filings that exceed certain thresholds for
magnitude of rate changes or filings that make use of new or sophisticated
predictive models may be subject to particular regulatory scrutiny. In these cases,
it is necessary to be able to explain the results of the modeling process in a
way that is understandable without sacrificing statistical rigor.

It should be noted that communicating results is not simply a method of passing
regulatory muster. Generating interpretable modeling output is an important -
even essential - facet of model checking. Actuaries are bound by relevant standards
to be able to exercise appropriate judgment in selecting risk characteristics as
part of a risk classification system per Actuarial Standard of Practice 12
(``Risk Classification''). Therefore, the techniques discussed
in this paper may be viewed from the lens of providing useful information to
regulators, but they should also be considered as part of a thorough vetting of
any rating model.

Although the focus of this paper is on communication to regulators, it should be
said that selecting final rates based on indications (step 4 in the list above) may
pose a unique challenge for black-box models. This, too, provides strong motivation
for techniques that could add to the modeler's - or any stakeholder's -
understanding of the model, such as the relative importance of variables or the
shapes of response curves. Such techniques could be usefully employed in making
decisions about how best to select rates.

Similarly, although the focus of this paper is on communication in a pricing
context, the techniques explored in this paper (and many of the concerns discussed)
may also be relevant to other contexts, such as claim-level reserving or analytics,
or other applications of machine learning to the insurance industry.

\hypertarget{need}{%
\section{The Need to See Inside the Black Box}\label{need}}

Within the actuarial profession, Actuarial Standard of Practice 41 (``Actuarial
Communications'') notes that ``\ldots another actuary qualified in the same practice
area {[}should be able to{]} make an objective appraisal of the reasonableness of
the actuary's work as presented in the actuarial report.'' (``Actuarial Standard of Practice No. 41 - Actuarial Communications'' 2010) Underlying
this requirement is an assumption that the hypothetical other actuary qualified
in the same practice area is adequately familiar with the relevant techniques
employed. Although the syllabus of basic education is constantly changing, there
has at times been an assumption that all techniques and assumptions that have
ever been a part of the syllabus of basic education needn't be explained from
first principles in general actuarial communications, and that an actuary
practicing in the same field should be able to make an objective appraisal of
the results from the methods found in the syllabus.
This is notable because, beginning with the introduction
of the CAS MAS-I and MAS-II examination in July of 2018, several machine learning
models were formally included in the syllabus of basic education. These exams
cover a wide range of topics, such as splines, clustering algorithms, decision
trees, boosting, and principle components analysis (Casualty Actuarial Society 2018).

Nevertheless, machine learning poses something of a special challenge for ASOP
41 for several reasons:

\begin{enumerate}
\def\labelenumi{\arabic{enumi}.}
\tightlist
\item
  Machine learning models can be very ad hoc compared to
  traditional statistical models.
\item
  Because many machine learning models do not assume an underlying probability
  distribution or stochastic process, they may not admit of standard metrics for model comparison
  (e.g., it's not straightforward to calculate an AIC over a neural network).
\item
  Machine learning methods are often combined into ensembles that
  may not be easily separated and that may, as a collection, cease to resemble
  a single standard version of a model.
\item
  Machine learning models can be ``black boxes'' insofar as the final form of
  response curve cannot be easily predicted and may depend heavily on the
  available data (which may not, in turn, be available to the reviewer).
\end{enumerate}

This last item raises a final interesting issue. GLMs and their ilk are often
fitted using one of a handful of standard and well-understood approaches (e.g.,
maximum likelihood estimation). However, this is not possible in general with
machine learning models, as machine learning algorithms often use loss surfaces
that are very complex such that it may not be feasible to calculate the global
minimum of the surface. Certainly, closed form representations of the loss
surfaces are not generally available. For this reason, the training phase of a
machine learning model is, in many ways, just as important to one's
understanding as the model form and the data on which the model is fitted.
Because the final model result is inseparable from these three components
(training method, model form, and data), it is not generally adequate to just
know the method employed to make an objective appraisal of the reasonableness
of the result.

These issues also pose particular challenges with respect to other standards.
For instance, as discussed previously, ASOP 12 requires
actuaries to be able to exercise appropriate judgment about risk classification
systems. The recent ASOP 56 (``Models'') speaks to more general concerns in all
practice areas that might make use of models. ASOP 56 requires the actuary to
``make reasonable efforts to confirm that the model structure, data, assumptions,
governance and controls, and model testing and output validation are consistent
with the intended purpose.'' (``Actuarial Standard of Practice No. 56 - Modeling'' 2019) All of these efforts may be hampered if
it is not possible to peer into the black box of the model.

It should also be noted that these comments only apply within the actuarial profession. Outside of
the actuarial profession, communication of results may be more challenging. A
2017 survey conducted by the Casualty Actuarial and Statistical Task Force of
the National Association of Insurance Commissioners (NAIC) found that the plurality of
responding regulators identified ``Filing complexity and/or a lack of resources
or expertise'' as a key challenge that impedes their ability to review GLMs or
other predictive models (National Association of Insurance Commissioners 2017). Given that machine learning
algorithms are generally regarded as more complex than GLMs, this implies that
the challenge of communicating machine learning model results is significant.

In response to the same survey, 33 state regulators noted that it would be
helpful or very helpful for the NAIC to develop information and tools to assist
in reviewing rate filings based on GLMs, and 34 noted that it would be helpful
to develop similar items to assist in reviewing ``Other Advanced Modeling
Techniques.'' One outgrowth of this need was the development of a white paper,
National Association of Insurance Commissioners (2019), on best practices for regulatory review of predictive models.
The white paper focuses on review of GLMs, particularly with respect to private
passenger automobile and homeowners' insurance. Some of the guidance offered in
this regard is therefore not strictly applicable to the review of machine
learning models. For example, as previously noted, p-values are not a concept
that translates well to deterministic machine learning algorithms. However,
among the guidance applicable to machine learning algorithms are the following:

\begin{itemize}
\tightlist
\item
  Determine the extent to which the model causes premium disruption for individual
  policyholders and how the insurer will
  explain the disruption to individual consumers that inquire about it.
\item
  Determine that individual input characteristics to a predictive model are
  related to the expected loss or expense
  differences in risk. Each input characteristic should have an intuitive or
  demonstrable actual relationship to expected loss or expense.
\item
  Determine that individual outputs from a predictive model and their associated
  selected relativities are not unfairly discriminatory.
\end{itemize}

The last of these items is an entire topic unto itself. The methods and
concepts introduced in this paper are useful for exploring the question of whether
rates are appropriately related to risk of loss as defined by the variables used
in the model, but there are many other aspects of discrimination-free that are
outside the scope of this paper. The methods in this paper may help in understanding
the model, which is a necessary precursor to addressing the question of unfair
discrimination.

The items in this list are by no means exhaustive, but they pertain to the concept of model
interpretability for ratemaking that we develop next.

\hypertarget{interpretability}{%
\section{Interpretability in the Ratemaking Context}\label{interpretability}}

In this section, we attempt to develop a working definition of interpretability
for ratemaking applications. While we will not provide a comprehensive survey of
the prolific and fast evolving ML interpretability literature, we draw from it
as appropriate in setting the stage for our discussion. Even among researchers
in the subject, there is not a consensus on the definition of interpretability;
here are a few from frequently cited papers:

\begin{itemize}
\tightlist
\item
  Ability to explain or to present in understandable terms to a human
  (Doshi-Velez and Kim 2017);
\item
  The degree to which an observer can understand the cause of a decision
  (Biran and Cotton 2017); and
\item
  A method is interpretable if a user can correctly and efficiently predict the
  method's results (Kim, Khanna, and Koyejo 2016).
\end{itemize}

We motivate our discussion by considering several aspects of interpretability.
As we proceed through the points below, we aim to arrive at a more scoped and
relevant definition of what it means for a pricing
model to be interpretable. In the remainder of this section, we clarify a couple concepts regarding
interpretable \emph{classes} of models and the computational transparency of ML
models, outline frameworks for understanding the communication goals of interpretability,
then discuss a potential framework for implementing ML
interpretability in practice.

\hypertarget{linear-models}{%
\subsection{Not All Linear Models are Interpretable}\label{linear-models}}

In the actuarial science literature, the GLM is probably the most oft-cited
example of an easily interpretable model. Given a set of inputs, we can easily
reason about what the output of the model is. As an illustrative example,
consider a claim severity model with driver age, sex, and vehicle age as
predictors; assuming a log link function and letting \(Y\) denote the response, we
have

\begin{align}
& \log(E[Y]) = \beta_0 + \beta_1 \cdot \text{age} \nonumber \\
& \quad + \beta_2 \cdot \text{vehicle\_age} + \beta_3 \cdot \text{sex}_{\text{male}}.
\end{align}

Here, we can tell, for example, what the model would predict for the expected
severity if we were to increase age by a certain amount, \emph{all else being equal},
because the relationship between the predictor and the response is simply
multiplication by the coefficient \(\beta_1\) and applying the inverse link
function.

\begin{figure}

{\centering \includegraphics[width=1\linewidth]{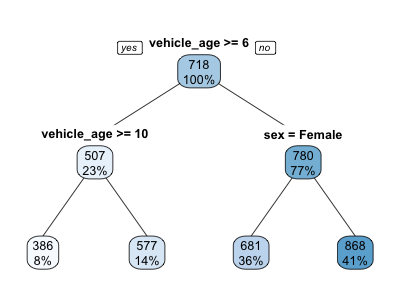} 

}

\caption{A simple decision tree for loss cost prediction.}\label{fig:tree-plot1}
\end{figure}

Another commonly cited example of an interpretable model is a decision tree. An
illustrative example is shown in Figure \ref{fig:tree-plot1}. Here, the
prediction is arrived at by following a sequence of if-else decisions.

Now, it is worth pointing out that, when declaring that GLMs or decision trees
are interpretable models, we are implicitly assuming that we are considering
only a handful of predictors. In fact, the ease with which we can reason about a
model declines as the number of predictors, transformations of them, and
interactions increase, as in the following (somewhat pathological) example:

\begin{align}\label{eq:badglm}
& \log(E[Y]) = \beta_0 + \beta_1 \cdot \text{age} + \beta_2 \cdot \text{vehicle\_age} \nonumber\\
& \quad + \beta_3\cdot\text{vehicle\_age}^2 \nonumber \\
& \quad + \beta_4 \cdot \text{age} \cdot \text{vehicle\_age} \nonumber \\
& \quad + \beta_5 \cdot \text{sex}_{\text{male}} + \beta_6 \cdot \text{sex}_{\text{male}} \cdot \text{age}. 
\end{align}

\begin{figure*}

{\centering \includegraphics[width=0.6\linewidth]{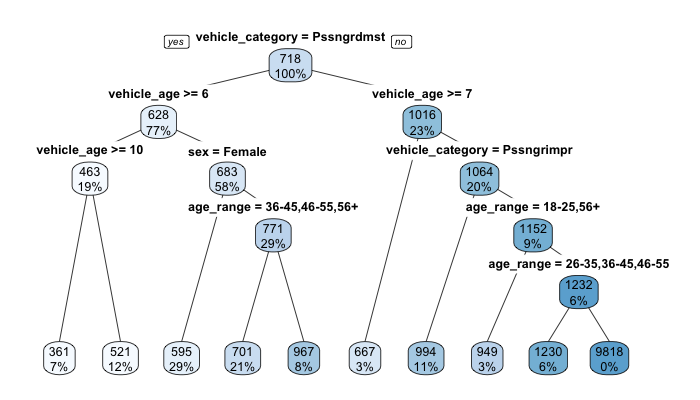} 

}

\caption{A more complex decision tree. This is still much simpler than typical realistic examples.}\label{fig:tree-plot2}
\end{figure*}

Similarly, one can see that in Figure \ref{fig:tree-plot2}, larger trees are
tough to reason about. In other words, even when working within the framework
of an ``interpretable'' class of models, we may still end up with something that
many would consider ``black box.''

\hypertarget{the-machinery-is-not-a-secret}{%
\subsection{The Machinery is Not a Secret}\label{the-machinery-is-not-a-secret}}

Another occasional misconception is that we have no visibility into how some ML
models \emph{compute} predictions, which renders them uninterpretable. Outside of
proprietary algorithms, all common ML models, including neural networks,
gradient boosted trees, and random forests, are well studied and have large
bodies of literature documenting their inner workings. As an example, a fitted
feedforward neural network is simply a composition of linear transformations
followed by nonlinear activation functions. As in Equation \ref{eq:badglm}, one
can write down the mathematical equation for calculating the prediction given
some inputs, but it may be difficult for a human to reason about it. We show
later that we can still provide explanations of completely ``black box'' models,
but is important to note that ML model predictions are still governed by
mathematical rules, and are deterministic in most cases.

\hypertarget{explanations-are-contextual}{%
\subsection{Explanations are Contextual}\label{explanations-are-contextual}}

Hilton (1990) proposed a framework, later interpreted by
Miller (2017) in the context of ML, for understanding model explanations
as \emph{conversations} or \emph{social interactions.} One consequence of this identification
is that explanations need to be \emph{relevant} to the \emph{audience}. This framework is
consistent with ASOP 41, which formulates a similar requirement in terms of an
\emph{intended user} of the actuarial communication. In developing,
filing, and operationalizing a pricing model, one needs to accommodate a variety
of stakeholders, each of whom has a different set of questions, assumptions,
and technical capacity. First, there are internal stakeholders at the company,
which includes management and underwriters. While some of the individuals in
this audience may be technical, they are likely less familiar with predictive
modeling techniques than the actuaries and data scientists who build the models.
Next, we have the regulators, who may have limited resources to review the
models, and will focus on a specific list of questions motivated by statute
and public policy. Finally, we have potential policyholders, who have an interest (perhaps
more so than the other parties) as they are responsible for paying the premiums.

It is interesting to note that the modelers, who are most familiar with the
models, tend to be same people designing and communicating the explanations.
This poses a challenge that Miller, Howe, and Sonenberg (2017) call ``inmates running
the asylum'', where the modelers design explanations for \emph{themselves} rather
than the intended audience. For example, they may be interested in technical
questions, such as extrapolation behavior, and shape the explanations accordingly,
which may be irrelevant to a prospective policyholder.

Another point outlined in Miller's survey (Miller 2017) is
that explanations are \emph{contrastive}. In other words, people are often interested
in not why something happened, but rather why it happened instead of something
else. For example, policyholders might not care exactly how their auto premiums
are computed, but would like to know why they are being charged more than their
coworkers who drive similar vehicles. As an extension, policyholders may want
to know what they can change in order to obtain lower premiums.

\hypertarget{asking-and-answering-the-right-questions}{%
\subsection{Asking and Answering the Right Questions}\label{asking-and-answering-the-right-questions}}

With the above considerations in mind, we propose a potential framework for
interpreting ML models for insurance pricing: the actuarial profession, in
collaboration with regulators and representatives of the public, define a set
of questions to be answered by explanations accompanying ML models, along
with acceptance criteria and examples of successful explanations. In other
words, interpretability for our purposes is defined as the ability of a model's
explanations to answer the posed questions.

It should be noted that no ideal set of questions exists that would encompass
all potential models. Rather, the actuary must consider what aspects of the
model would raise questions from the perspective of the model's intended users.
We propose that relevant stakeholders, by providing example questions and
answers, would inherently provide guidance by which actuaries can reasonably
anticipate the kinds of specific questions most important to those stakeholders
and address them proactively.

These questions should relate to existing guidelines, such as those described in
National Association of Insurance Commissioners (2019) and outlined in Section \ref{need}, standards of practice, and
regulation, and in fact should not be specific only to ML models. By
conceptualizing a set of questions, we reduce the burden of both companies and
regulators; this is especially important for the latter, who are already
resource constrained facing increasing variety of models being filed. This
format should also be familiar to actuaries who are accustomed to adhering to
specific guidelines in, for example, ASOPs. Like the ASOPs, We envision that
these questions and guidelines will be continually updated to reflect feedback
obtained and advances in research.

While the realization of a set of such guidelines is an ambitious undertaking
beyond the scope of this paper, we present in the next section a sample set of
questions and techniques one can leverage to answer them. The goal of these case
studies is twofold: to more concretely illustrate the proposed framework, and
to expose the actuarial audience to modern ML interpretation techniques.

\hypertarget{application}{%
\section{Applying Model Interpretation Techniques}\label{application}}

Now that we have established a framework for model interpretation in the form of
asking and answering relevant questions, we demonstrate examples of such
exchanges via an illustrative case study. Analytically, our starting point is a
fitted deep neural network model for predicting loss costs. As the modeling
details are of secondary importance, they are available in Appendix
\ref{model-dev}. The questions that we ask of the model are as follows:

\begin{enumerate}
\def\labelenumi{\arabic{enumi}.}
\tightlist
\item
  What are the most important predictors in the model? Put another way, to
  what extent do the predictors improve the accuracy of the model?
\item
  How does the predicted loss cost change, on average, as we change an input?
\item
  For a particular policyholder, how does each characteristic contribute to
  the loss cost prediction?
\end{enumerate}

The techniques we utilize to answer these questions are permutation variable
importance, partial dependence plots, and additive variable attributions,
respectively.
In our discussion, we adopt the organization of techniques and some notation
presented in Molnar and others (2018) and Biecek and Burzykowski (2019), which are comprehensive
references on the most established ML interpretation techniques.

\hypertarget{a-simplified-view-of-interpretation-techniques}{%
\subsection{A Simplified View of Interpretation Techniques}\label{a-simplified-view-of-interpretation-techniques}}

Before we dive into the answering questions, we present a brief taxonomy of ML
interpretation techniques. Rather than attempting an exhaustive classification,
the goal is to orient ourselves among broad categories of techniques, so we can
map them to tasks indicated by the questions being asked. For our purposes,
model interpretation techniques can be categorized across two dimensions:
intrinsic vs.~post-hoc and global vs.~local.

\hypertarget{intrinsic-vs.-post-hoc}{%
\subsubsection{Intrinsic vs.~Post-hoc}\label{intrinsic-vs.-post-hoc}}

Intrinsic model interpretation draws conclusions from the structure of the
fitted model and are what we typically associate with ``interpretable'' classes
of models. This is only viable with models with simple structures, such as the
sparse linear model and shallow decision tree we see in Section
\ref{linear-models}, where we arrive at explanations by reading off parameter
estimates or a few decision rules. For algorithms that produce models
with complex structure that do not lend themselves easily to intrinsic
exploration, we can appeal to post-hoc techniques. This class of techniques
interrogate the model by presenting it with data for scoring and observing the
prediction behavior of the model. These techniques are concerned with only the
inputs and outputs, and hence are \emph{agnostic} of the model itself, which means
they can also be applied to simple models. Since most useful ML models have a
level of complexity beyond the threshold of intrinsic interpretability, we focus
on model-agnostic techniques in our case study. As we will see later on, the
data that we present to the models are usually some perturbed variations of test
data.

\hypertarget{global-vs.-local}{%
\subsubsection{Global vs.~Local}\label{global-vs.-local}}

Along the other dimension, we categorize model interpretations as global, or
model-level, and local, or instance-level. The former class provides insights
with respect to the model as a whole. Some examples of these eplanations include
variable importances and sensitivities, on average, of the predicted response
with respect to individual predictors. Note that these methods may be compared to
the methods described in (Goldburd, Khare, and Tevet 2016), Chapter 7, which focus on global
interpretation of GLMs.

In our case study, questions 1 and 2 are
associated with global interpretations. On the other hand, question 3 pertains to an
individual prediction, which would fall in the local, or instance-level,
category. In addition to individual variable attribution, we can also inquire
about what would happen to the current predicted response if we were to perturb
specific predictor variables.

\hypertarget{answering-the-questions}{%
\subsection{Answering the Questions}\label{answering-the-questions}}

Having aligned the questions with the categories of interpretation techniques,
we now introduce a selection of appropriate techniques to answer them.

\hypertarget{variable-importance}{%
\subsubsection{Variable Importance}\label{variable-importance}}

``What are the most important predictors in the model?''

For linear models and their generalizations, and some ML models, measures of
variable importance can be obtained from the fitted model structure. In the
case of GLMs, one might observe the magnitudes of the estimated coefficients or
\(t\)-statistics, whereas
for random forests, one might use out-of-bag errors (Breiman 2001).
For more complex models, such as the neural network in our case study, we need
to devise another approach.

We follow the methodology of permutation feature importance as described in
Fisher, Rudin, and Dominici (2018), and utilize the notation introduced by Biecek and Burzykowski (2019).
The gist of the technique is as follows: to see how important a variable is,
we make predictions without it and see how much worse off we are in terms of
accuracy. One way to achieve this would be to re-fit the model many times (as
many times as the number of variables.) However, this may be intractable with
lengthy model training times or large numbers of variables, so a more popular
approach is to instead keep the same fitted model but permute the values of each
predictor.

More formally, let \(y\) denote the vector of responses, \(X\) denote the matrix of
predictor variables, \(\widehat{f}\) denote the fitted model, and
\(L = \mathcal{L}(\widehat{f}(X), y)\), where \(\widehat{f}\) applies to \(X\)
rowwise, denote the value of the loss function, which is mean squared error
in the case of regression. Now, if \(\widetilde{X}^{j}\) denotes the predictor
matrix where the \(j\)th variable has been permuted, then we can compute the
loss with the permuted dataset as
\(L^{-j} = \mathcal{L}(\widehat{f}(\widetilde{X}^{j}), y)\).
Here, by \emph{permuting} a variable, we mean that we randomly rearrange the values
in the column of data associated with the variable.
With this, we define
the variable importance \(VI^{j}\) as \(L^{-j} - L\).

\begin{figure}

{\centering \includegraphics[width=1\linewidth]{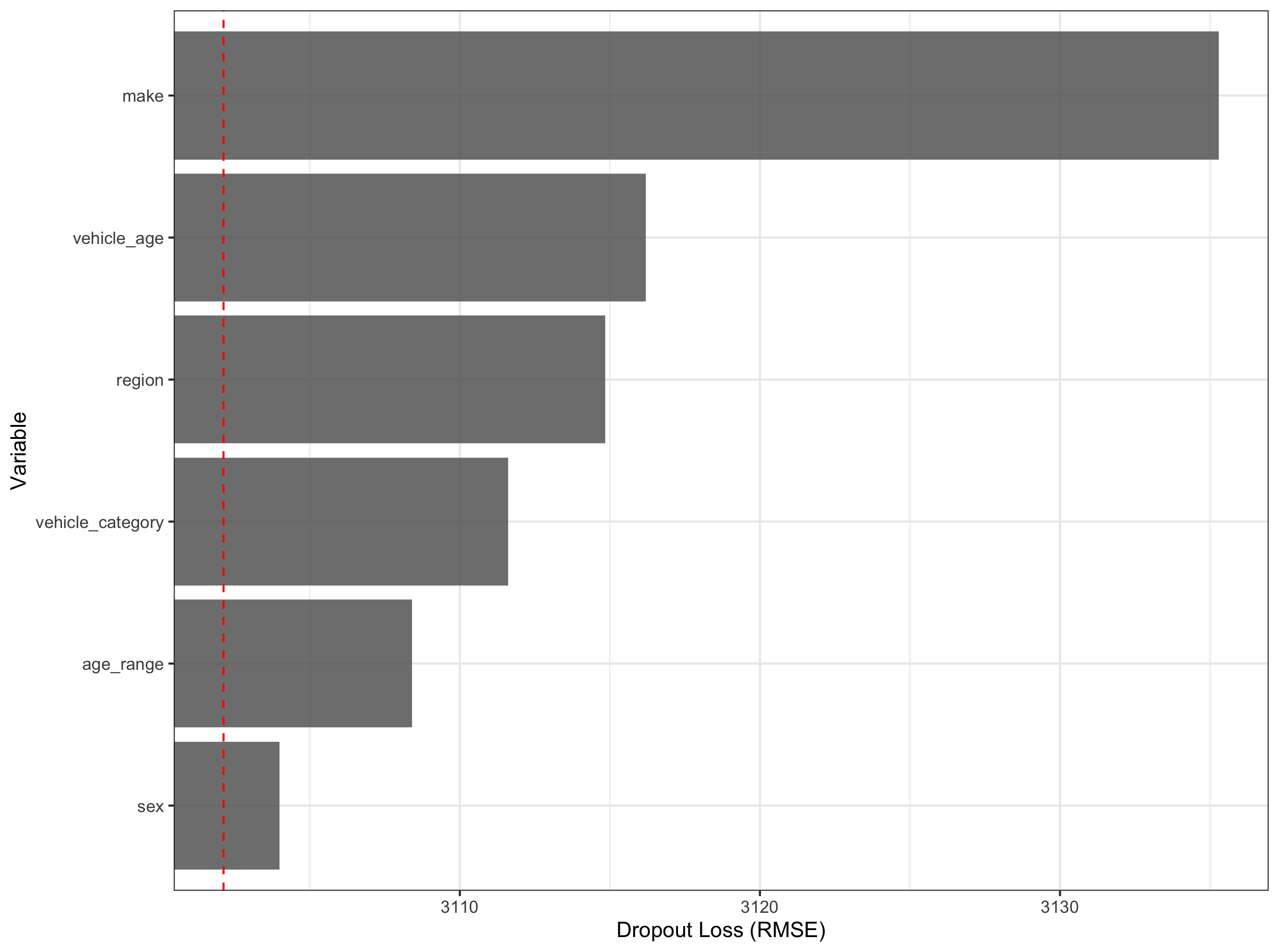} 

}

\caption{Permutation feature importances for the neural network model.}\label{fig:fi}
\end{figure}

In Figure \ref{fig:fi}, we show a plot of variable importances. In our
particular example, we see that the ``make'' variable contributes most to the
accuracy of the model with ``sex'' contributing the least. This provides a way
for the audience to quickly glance at the most relevant variables, and ask
further question as necessary.

Note that these measures do not provide information regarding the directional
sensitivity of the predictors on the response. Also, when there are correlated
variables, one should be careful about interpretation, as the result may be
biased by unrealistic records in the permuted dataset. Another ramification
of a group of correlated variables is that their inclusion may cause each to
appear less important than if only one is included in the model.

\hypertarget{partial-dependence-plots}{%
\subsubsection{Partial Dependence Plots}\label{partial-dependence-plots}}

``How does the predicted loss cost change, on average, as we change an input?''

For this question, we again consider first how it would be answered in the GLM
setting. When the input predictor in question is continuous, we can answer the
question by looking at the estimated coefficient, which provides the change in
the response per unit change in the predictor (on the scale of the linear
predictor). For non-parametric models and neural networks, where no coefficients
are available, we can appeal to partial dependence plots (PDP), first proposed
by Friedman (2001) for gradient boosting machines (GBM).

\begin{figure*}[h]

{\centering \includegraphics[width=0.7\linewidth]{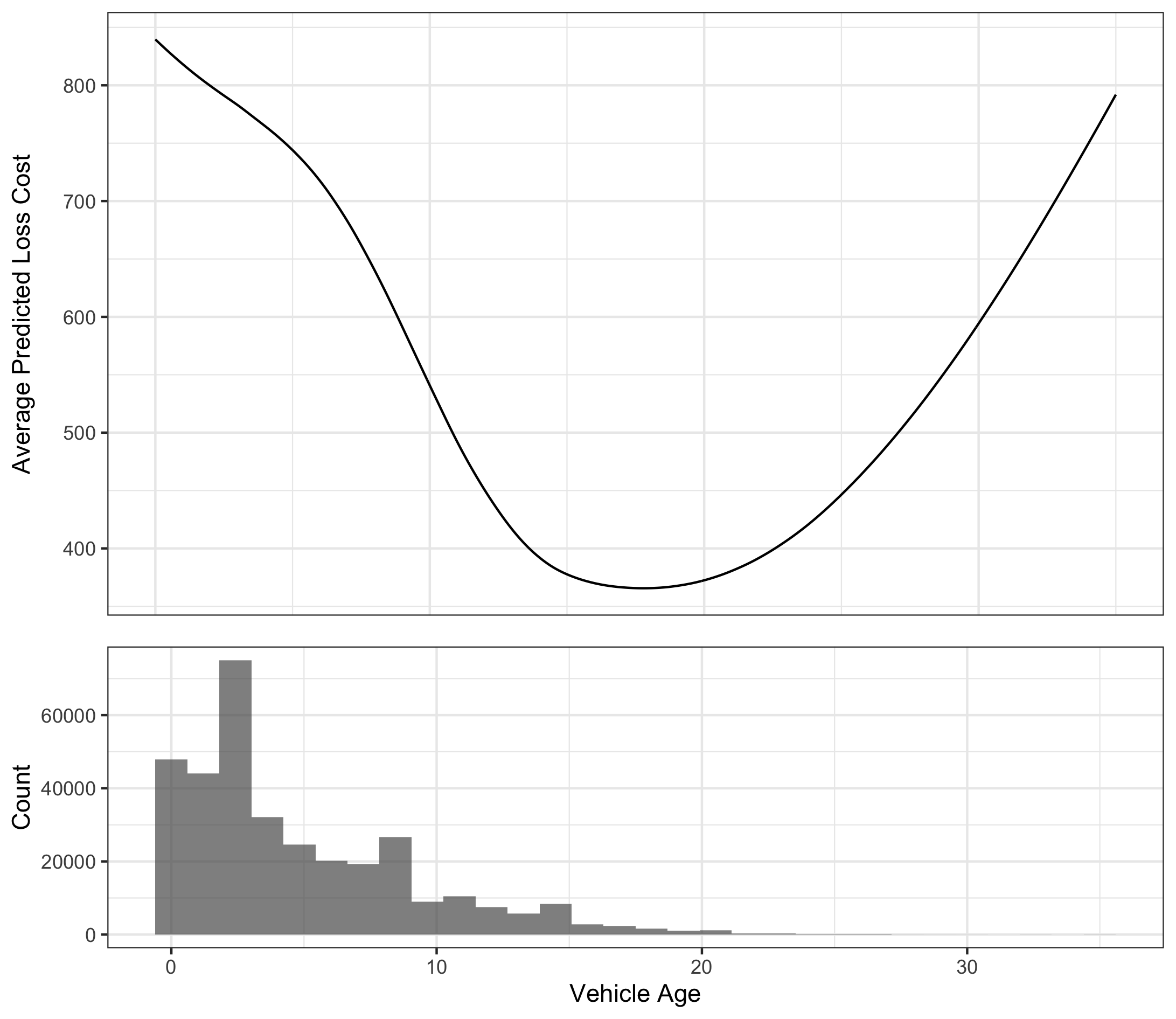} 

}

\caption{Partial dependence plot for the neural network model.}\label{fig:pdp}
\end{figure*}

\begin{figure*}[h]

{\centering \includegraphics[width=0.8\linewidth]{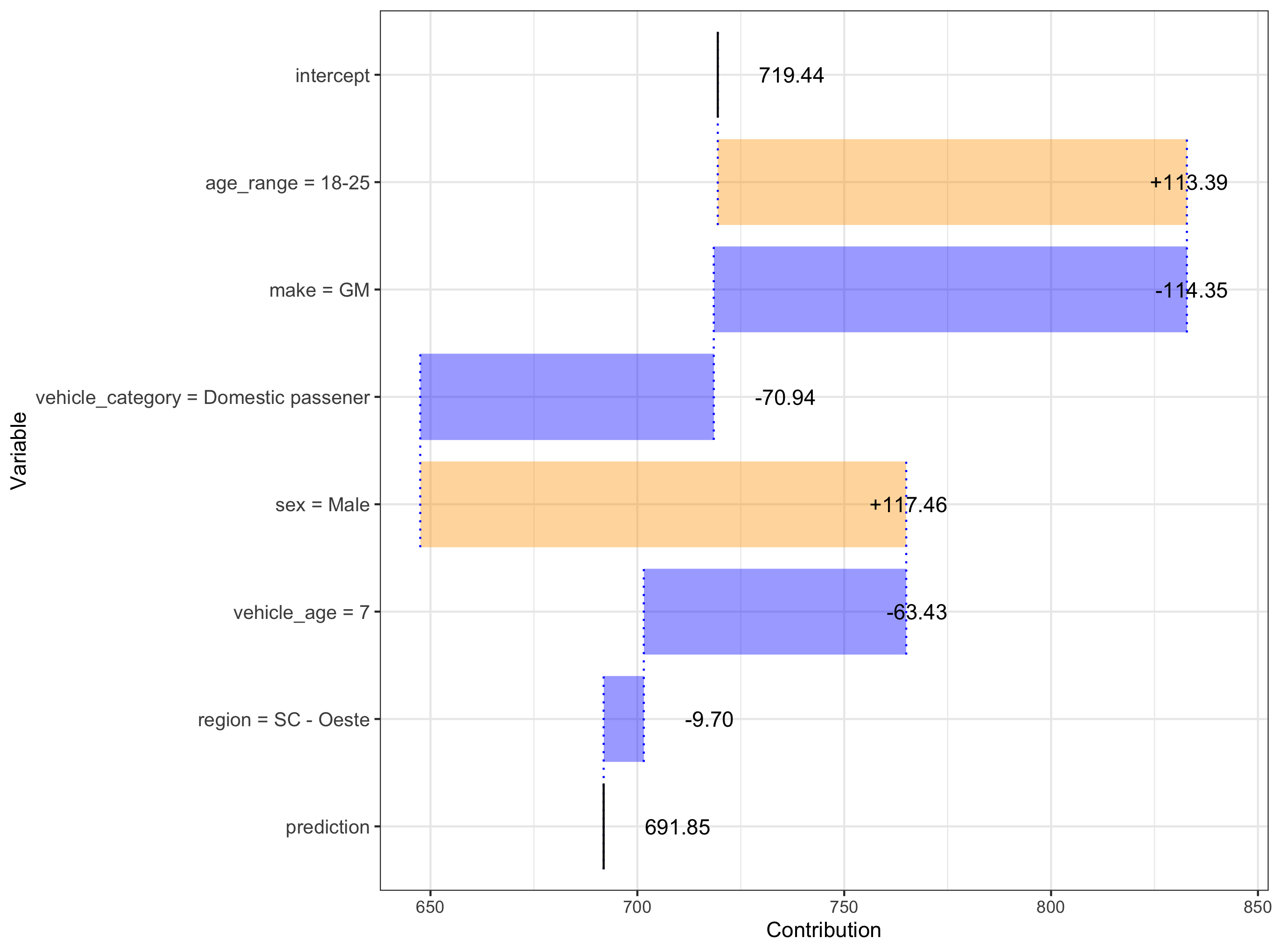} 

}

\caption{Variable contribution plot for the neural network model.}\label{fig:breakdown}
\end{figure*}

To describe PDP, we need to introduce some additional notation. Let \(x^j\) denote
the input variable of interest. Then we define the partial dependence function as

\begin{equation}
h(z) = E_{X^{-j}}[\widehat{f}(x|x^j = z)],
\end{equation}

where the expectation is taken over the distribution of the other predictor
variables. In other words, we marginalize them out so we can focus on the
relationship between the predicted response and the variable of interest.
Empirically, we estimate \(h\) by

\begin{equation}
\widehat{h}(z) = \frac{1}{N}\sum_{i = 1}^{N}\widehat{f}(x_i|x_i^{j} = z),
\end{equation}

where \(N\) is the number of records in the dataset.

In Figure \ref{fig:pdp}, we exhibit the PDP for the ``vehicle age'' variable.
We see that the average predicted loss cost decreases with vehicle age until
the latter is around 18. Note that the accompanying histogram shows that the
data is quite thin for vehicle age greater than 18, so the apparent upward
trend to the right is driven by just a few data points.

This information allows the modeler and stakeholders to consider whether it
is reasonable for the anticipated loss cost to follow this shape.

The question posed here is particularly important for regulators, who would
like to know whether each variable affects the prediction in the direction that
is expected, based on intuition, experience, and existing models. During the
model development stage, PDP can also be used as a reasonableness test for
candidate models by identifying unexpected relationships for the analyst to
investigate.

As with permutation feature importance, one should be careful when interpreting
PDP when there are strongly correlated variables. Since we average over the
marginal distribution of the rest of the variables, we may take into account
unrealistic data (e.g.~high vehicle age for a model that is brand new). To
address this drawback, alternative visualization techniques have been proposed,
such as accumulated local effect (ALE) plots, which take expectations over
conditional, rather than marginal, distributions (Apley and Zhu 2016).

\hypertarget{variable-attribution}{%
\subsubsection{Variable Attribution}\label{variable-attribution}}

``For a particular policyholder, how does each characteristic contribute to the
loss cost prediction?''

In the previous two examples, we look at model-level explanations; now we
move on to one where we investigate one particular prediction instance. As
before, we consider how we would approach the question for linear models. For
a GLM with a log link common in ratemaking applications, for example, we start
with the base rate, then the exponentiated coefficients would have
multiplicative effects on the final rate. Similar to the previous examples,
for ML models in general we do not have directly interpretable weights.
Instead, one way to arrive at variable contributions is calculating the change
in the expected model prediction for each predictor, conditioned on other
predictors.

Formally, for a fitted model \(\widehat{f}\), a given ordering of the variables
\(X^1,\dots,X^p\), where \(p\) is the number of predictor variables,
and a specific instance \(x_*\), we
would like to decompose the model prediction \(\widehat{f}(x_*)\) into

\begin{equation}\label{eq:contrib-decompose}
\widehat{f}(x_* ) = v_0 + \sum_{j=1}^p v(j, x_*),
\end{equation}

where \(v_0\) denotes the average model response, and \(v(j, x_*)\) denotes the
contribution of the \(j\)th variable in instance \(x_*\), defined as

\begin{align}\label{eq:var-contrib}
& v(j, x_* ) = E_X[\widehat{f}(X) | X^1 = x_* ^1, \dots, X^j = x_* ^j] \nonumber\\
& \quad - E_X[\widehat{f}(X) | X^1 = x_* ^1, \dots, X^{j-1} = x_*^{j - 1})].
\end{align}

Hence, the contribution of the \(j\)th variable to the prediction is the
incremental change in the expected model prediction when we set \(X^j = x_*^ j\)
assuming the other variables take their values in \(x_*\). Note here that this
definition implies that the order in which we consider the variables affects
the results. Empirically, the expectations in \eqref{eq:var-contrib} are
calculated by sampling the test dataset.

In Figure \ref{fig:breakdown}, we exhibit a waterfall plot of variable
contributions. The ``intercept'' value denotes the average model prediction and
represents the \(v_0\) term in Equation \eqref{eq:contrib-decompose}. The predicted
loss cost for this particular policyholder is slightly less than average; the
characteristics that makes this policyholder more risky are the fact that he is
a male between the ages of 18 and 25; counteracting the risky driver
characteristics are the vehicle properties: it is a GM vehicle built
domestically and is seven years old.

Instance-level explanations are useful for investigating specific problematic
predictions generated by the model. Regulators and model reviewers may be
interested in variable contributions for the safest and riskiest policyholders
to see if they conform to intuition. A policyholder with a particularly high
premium may wish to find out what of their characteristics contribute to it,
and may follow up with a question about how he can lower it, which would
require another type of explanation.

As noted earlier, the ordering of variables has an impact on the contributions
calculated, especially for models that are non-additive, which could cause
inconsistent explanations. There are several approaches to ameliorate this
phenomenon, including selecting variables with the largest contributions first,
including interactions terms, and averaging over possible orderings. The last
of these ideas is implemented by Lundberg and Lee (2017) using Shapley values from
cooperative game theory, and is referred to as Shapley additive explanations
(SHAP). These approaches are discussed further in Biecek and Burzykowski (2019) and its references.
Shapley values were also used in Mango (1998), which has appeared on the CAS syllabus
starting in 2004, in the context of determining how to allocate catastrophe
risk loads between multiple accounts.

\hypertarget{other-techniques}{%
\subsection{Other Techniques}\label{other-techniques}}

In this paper, we demonstrate just a few model-agnostic ML interpretation
techniques. These represent a small subset of existing techniques, each of
which have additional variations. In the remainder of this section, we point
out a few common techniques not covered in our case study.

Individual conditional expectation (ICE) plots disaggregate PDPs into their
instance-level components for a more granular view into predictor sensitivities
(Goldstein et al. 2015). To accommodate correlated variables in PDP, accumulated
local effect (ALE) plots computes expected changes in model response over the
conditional, rather than marginal, distribution of the other variables
(Apley and Zhu 2016).

Local interpretable model-agnostic explanations (LIME) (Ribeiro, Singh, and Guestrin 2016)
builds simple surrogate models using model predictions, with higher training
weights given to the point of interest, in effect replacing the complex ML model
with an easily interpretable linear regressions or decision trees in
neighborhoods of specific points for the purpose of explanation. Taking the
concept further, one can also train a global surrogate model across the entire
domain of interest.

\hypertarget{conclusion}{%
\section{Conclusion}\label{conclusion}}

Actuarial standards of practice, most notably ASOP 41, places responsibility on
the actuary to clearly communicate actuarial work products, including insurance
pricing models. These responsibilities create special challenges for
communicating machine learning models, which are often seen ``black boxes'' due
in part to their complexity, nonlinearity, flexible construction, and ad hoc
nature.

In this paper, we discuss particular questions of model validation that are
of key importance in communicating a model that may present particular
difficulty for machine learning models compared to GLMs or traditional pricing
models. Specifically,

\begin{itemize}
\tightlist
\item
  How does the model impact individual insurance consumers?
\item
  How are the predictor variables related to expected losses?
\end{itemize}

We contextualize these questions in terms of different frameworks for defining
interpretability. We conceptualize interpretability in terms of the ability of
a model (or modeler) to answer a set of idealized questions that would be
refined. We then offer potential (families of) model-agnostic techniques for
providing answers to these questions.

Much work remains to be done in terms of defining the role of machine learning
algorithms in actuarial practice. Lack of interpretability has been a key
barrier preventing wider adoption and exploration of these techniques. The
methods proposed in this paper could therefore represent important strides in
unlocking the potential of machine learning within the insurance industry.

\hypertarget{acknowledgments}{%
\section*{Acknowledgments}\label{acknowledgments}}
\addcontentsline{toc}{section}{Acknowledgments}

We thank Navdeep Gill, Daniel Falbel, Morgan Bugbee, the volunteers of the CAS
project oversight group, and two anonymous reviewers for helpful discussions
and/or feedback.
This work is supported by the Casualty Actuarial Society.

\hypertarget{references}{%
\section*{References}\label{references}}
\addcontentsline{toc}{section}{References}

\hypertarget{refs}{}
\leavevmode\hypertarget{ref-tensorflow2015-whitepaper}{}%
Abadi, Martin, Ashish Agarwal, Paul Barham, Eugene Brevdo, Zhifeng Chen, Craig Citro, Greg S. Corrado, et al. 2015. ``TensorFlow: Large-Scale Machine Learning on Heterogeneous Systems.'' \url{http://tensorflow.org/}.

\leavevmode\hypertarget{ref-asop_41}{}%
``Actuarial Standard of Practice No. 41 - Actuarial Communications.'' 2010. \url{http://www.actuarialstandardsboard.org/wp-content/uploads/2014/02/asop041_120.pdf}.

\leavevmode\hypertarget{ref-asop_56}{}%
``Actuarial Standard of Practice No. 56 - Modeling.'' 2019. \url{http://www.actuarialstandardsboard.org/wp-content/uploads/2020/01/asop056_195.pdf}.

\leavevmode\hypertarget{ref-anderson_2005}{}%
Anderson, D., S. Feldblum, C. Modlin, D. Schirmacher, E. Schirmacher, and N. Thandi. 2005. ``A Practitioner's Guide to Generalized Linear Models,'' 4--39.

\leavevmode\hypertarget{ref-apley2016visualizing}{}%
Apley, Daniel W., and Jingyu Zhu. 2016. ``Visualizing the Effects of Predictor Variables in Black Box Supervised Learning Models.'' \url{http://arxiv.org/abs/1612.08468}.

\leavevmode\hypertarget{ref-bailey_simon_1960}{}%
Bailey, Robert A., and LeRoy J. Simon. 1960. ``Two Studies in Automobile Insurance Ratemaking.'' \emph{ASTIN Bulletin} 1 (4): 192--217. \url{https://doi.org/10.1017/S0515036100009569}.

\leavevmode\hypertarget{ref-biecekPMVEE}{}%
Biecek, Przemyslaw, and Tomasz Burzykowski. 2019. ``Predictive Models: Explore, Explain, and Debug.'' \url{https://pbiecek.github.io/PM_VEE/}.

\leavevmode\hypertarget{ref-biranExplanationJustification2017}{}%
Biran, Or, and Courtenay Cotton. 2017. ``Explanation and Justification in Machine Learning: A Survey.'' In \emph{IJCAI-17 Workshop on Explainable AI (XAI)}, 8:1.

\leavevmode\hypertarget{ref-breimanRandomForests2001}{}%
Breiman, Leo. 2001. ``Random Forests.'' \emph{Machine Learning} 45 (1): 5--32.

\leavevmode\hypertarget{ref-brown_1988}{}%
Brown, Robert L. 1988. ``Minimum Bias with Generalized Linear Models,'' 187--217.

\leavevmode\hypertarget{ref-cas_syllabus_2018}{}%
Casualty Actuarial Society. 2018. ``Syllabus of Basic Education'' 2018. \url{https://www.casact.org/admissions/syllabus/ArchivedSyllabi/2018Syllabus.pdf}.

\leavevmode\hypertarget{ref-doshi-velezRigorousScience2017}{}%
Doshi-Velez, Finale, and Been Kim. 2017. ``Towards A Rigorous Science of Interpretable Machine Learning.'' \emph{arXiv:1702.08608 {[}Cs, Stat{]}}, February. \url{http://arxiv.org/abs/1702.08608}.

\leavevmode\hypertarget{ref-dugas_2003}{}%
Dugas, Charles, Y. Bengio, Nicolas Chapados, P. Vincent, G. Denoncourt, and C. Fournier. 2003. ``Statistical Learning Algorithms Applied to Automobile Insurance Ratemaking,'' December. \url{https://doi.org/10.1142/9789812794246_0004}.

\leavevmode\hypertarget{ref-fisherAllModels2018}{}%
Fisher, Aaron, Cynthia Rudin, and Francesca Dominici. 2018. ``All Models Are Wrong, but Many Are Useful: Learning a Variable's Importance by Studying an Entire Class of Prediction Models Simultaneously.'' \emph{arXiv:1801.01489 {[}Stat{]}}, January. \url{http://arxiv.org/abs/1801.01489}.

\leavevmode\hypertarget{ref-friedmanGreedyFunction2001}{}%
Friedman, Jerome H. 2001. ``Greedy Function Approximation: A Gradient Boosting Machine.'' \emph{Annals of Statistics}, 1189--1232.

\leavevmode\hypertarget{ref-gabrielliNeuralNetwork2019}{}%
Gabrielli, Andrea. 2019. ``A Neural Network Boosted Double over-Dispersed Poisson Claims Reserving Model.'' SSRN Scholarly Paper ID 3365517. Rochester, NY: Social Science Research Network.

\leavevmode\hypertarget{ref-gabrielliNeuralNetwork2019a}{}%
Gabrielli, Andrea, Ronald Richman, and Mario V. Wüthrich. 2019. ``Neural Network Embedding of the over-Dispersed Poisson Reserving Model.'' \emph{Scandinavian Actuarial Journal}, 1--29.

\leavevmode\hypertarget{ref-gao_2018}{}%
Gao, Guangyuan, Shengwang Meng, and Mario V. Wuthrich. 2018. ``Claims Frequency Modeling Using Telematics Car Driving Data.'' \emph{Scandinavian Actuarial Journal}.

\leavevmode\hypertarget{ref-gao_2019}{}%
Gao, Guangyuan, and Mario Wuthrich. 2019. ``Convolutional Neural Network Classification of Telematics Car Driving Data.'' \emph{Risks} 7 (January): 6. \url{https://doi.org/10.3390/risks7010006}.

\leavevmode\hypertarget{ref-gao_2018_2}{}%
Gao, Guangyuan, and Mario V. Wuthrich. 2018. ``Feature Extraction from Telematics Car Driving Heatmaps.'' \emph{European Actuarial Journal} 8 (2): 383--406.

\leavevmode\hypertarget{ref-goldburd_2016}{}%
Goldburd, Mark, Anand Khare, and Dan Tevet. 2016. ``Generalized Linear Models for Insurance Rating.'' \emph{Casualty Actuarial Society, CAS Monographs Series}, no. 5.

\leavevmode\hypertarget{ref-goldstein2015peeking}{}%
Goldstein, Alex, Adam Kapelner, Justin Bleich, and Emil Pitkin. 2015. ``Peeking Inside the Black Box: Visualizing Statistical Learning with Plots of Individual Conditional Expectation.'' \emph{Journal of Computational and Graphical Statistics} 24 (1): 44--65.

\leavevmode\hypertarget{ref-henckaerts_2018}{}%
Henckaerts, Roel, Katrien Antonio, Maxime Clijsters, and Roel Verbelen. 2018. ``A Data Driven Binning Strategy for the Construction of Insurance Tariff Classes.'' \emph{Scandinavian Actuarial Journal} 2018 (8): 681--705.

\leavevmode\hypertarget{ref-hiltonConversationalProcesses1990}{}%
Hilton, Denis J. 1990. ``Conversational Processes and Causal Explanation.'' \emph{Psychological Bulletin} 107 (1): 65.

\leavevmode\hypertarget{ref-kimExamplesAre2016}{}%
Kim, Been, Rajiv Khanna, and Oluwasanmi O. Koyejo. 2016. ``Examples Are Not Enough, Learn to Criticize! Criticism for Interpretability.'' In \emph{Advances in Neural Information Processing Systems}, 2280--8.

\leavevmode\hypertarget{ref-kuoDeepTriangleDeep2018}{}%
Kuo, Kevin. 2019. ``DeepTriangle: A Deep Learning Approach to Loss Reserving.'' \emph{Risks} 7 (3): 97.

\leavevmode\hypertarget{ref-lange_1966}{}%
Lange, Jeffrey T. 1966. ``General Liability Insurance Ratemaking.'' \emph{Proceedings of the Casualty Actuarial Society} LIII: 26--53.

\leavevmode\hypertarget{ref-lecunDeepLearning2015}{}%
LeCun, Yann, Yoshua Bengio, and Geoffrey Hinton. 2015. ``Deep Learning.'' \emph{Nature} 521 (7553): 436.

\leavevmode\hypertarget{ref-NIPS2017_7062}{}%
Lundberg, Scott M, and Su-In Lee. 2017. ``A Unified Approach to Interpreting Model Predictions.'' In \emph{Advances in Neural Information Processing Systems 30}, edited by I. Guyon, U. V. Luxburg, S. Bengio, H. Wallach, R. Fergus, S. Vishwanathan, and R. Garnett, 4765--74. Curran Associates, Inc. \url{http://papers.nips.cc/paper/7062-a-unified-approach-to-interpreting-model-predictions.pdf}.

\leavevmode\hypertarget{ref-mango_1998}{}%
Mango, Donald F. 1998. ``An Application of Game Theory: Property Catastrophe Risk Load.'' In \emph{Proceedings of the Casualty Actuarial Society}, LXXXV:157--86.

\leavevmode\hypertarget{ref-mildenhall_1999}{}%
Mildenhall, Stephen J. 1999. ``A Systematic Relationship Between Minimum Bias and Generalized Linear Models.'' \emph{Proceedings of the Casualty Actuarial Society} LXXXVI: 393--487.

\leavevmode\hypertarget{ref-millerExplanationArtificial2017}{}%
Miller, Tim. 2017. ``Explanation in Artificial Intelligence: Insights from the Social Sciences.'' \emph{arXiv:1706.07269 {[}Cs{]}}, June. \url{http://arxiv.org/abs/1706.07269}.

\leavevmode\hypertarget{ref-millerExplainableAI2017}{}%
Miller, Tim, Piers Howe, and Liz Sonenberg. 2017. ``Explainable AI: Beware of Inmates Running the Asylum or: How I Learnt to Stop Worrying and Love the Social and Behavioural Sciences.'' \emph{arXiv Preprint arXiv:1712.00547}.

\leavevmode\hypertarget{ref-molnar2018interpretable}{}%
Molnar, Christoph, and others. 2018. ``Interpretable Machine Learning: A Guide for Making Black Box Models Explainable.'' \emph{E-Book At\textless{} Https://Christophm. Github. Io/Interpretable-Ml-Book/\textgreater, Version Dated} 10.

\leavevmode\hypertarget{ref-naic_summer_2017}{}%
National Association of Insurance Commissioners. 2017. ``2017 Proceedings of the National Association of Insurance Commissioners.''

\leavevmode\hypertarget{ref-naic_white_paper}{}%
---------. 2019. ``Regulatory Review of Predictive Models 10/15/19 Exposure Draft.'' \url{https://content.naic.org/sites/default/files/inline-files/Predictive\%20Model\%20White\%20Paper\%20Exposed\%2010-15-19.docx}.

\leavevmode\hypertarget{ref-nelder_wedderburn_1972}{}%
Nelder, J. A., and R. W. M. Wedderburn. 1972. ``Generalized Linear Models.'' \emph{Journal of the Royal Statistical Society}, Series a (general), 135 (3): 370--84.

\leavevmode\hypertarget{ref-nilsson_2009}{}%
Nilsson, Nils J. 2009. \emph{The Quest for Artificial Intelligence}. Cambridge University Press.

\leavevmode\hypertarget{ref-noll_2018}{}%
Noll, Alexander, Robert Salzmann, and Mario Wuthrich. 2018. ``Case Study: French Motor Third-Party Liability Claims.'' \emph{SSRN Electronic Journal}.

\leavevmode\hypertarget{ref-rlang}{}%
R Core Team. 2019. \emph{R: A Language and Environment for Statistical Computing}. Vienna, Austria: R Foundation for Statistical Computing. \url{https://www.R-project.org/}.

\leavevmode\hypertarget{ref-ribeiro2016should}{}%
Ribeiro, Marco Tulio, Sameer Singh, and Carlos Guestrin. 2016. ``Why Should I Trust You?: Explaining the Predictions of Any Classifier.'' In \emph{Proceedings of the 22nd Acm Sigkdd International Conference on Knowledge Discovery and Data Mining}, 1135--44. ACM.

\leavevmode\hypertarget{ref-robertson_2009}{}%
Robertson, J. P. 2009. ``NCCI's 2007 Hazard Group Mapping.'' \emph{Variance} 3 (2): 194--213.

\leavevmode\hypertarget{ref-roel_2018}{}%
Roel, Verbelen, Katrien Antonio, and Gerda Claeskens. 2018. ``Unraveling the Predictive Power of Telematics Data in Car Insurance Pricing.'' \emph{Royal Statistical Society}.

\leavevmode\hypertarget{ref-spedicato_2018}{}%
Spedicato, Giorgio, Christophe Dutang, and Leonardo Petrini. 2018. ``Machine Learning Methods to Perform Pricing Optimization: A Comparison with Standard Generalized Linear Models.'' \emph{Variance} 12 (1).

\leavevmode\hypertarget{ref-wang_raj_2017}{}%
Wang, Haohan, and Bhiksha Raj. 2017. ``On the Origin of Deep Learning.'' \emph{arXiv:1702.07800v4 {[}cs.LG{]}}.

\leavevmode\hypertarget{ref-wuthrich_2017}{}%
Wuthrich, Mario V. 2017. ``Covariate Selection from Telematics Car Driving Data.'' \emph{European Actuarial Journal} 7 (1): 89--108.

\leavevmode\hypertarget{ref-wuthrichMachineLearning2018}{}%
Wüthrich, Mario V. 2018. ``Machine Learning in Individual Claims Reserving.'' \emph{Scandinavian Actuarial Journal} 2018 (6): 465--80.

\leavevmode\hypertarget{ref-yang_2018}{}%
Yang, Yi, Wei Qian, and Hui Zou. 2018. ``Insurance Premium Prediction via Gradient Tree-Boosted Tweedie Compound Poisson Models.'' \emph{Journal of Business and Economic Statistics} 36 (3): 456--70.

\appendix

\hypertarget{appendix}{%
\section*{Appendix}\label{appendix}}
\addcontentsline{toc}{section}{Appendix}

\begin{table}[!h]

\caption{\label{tab:variables}Input variables and their transformations.}
\centering
\begin{tabular}[t]{lll}
\toprule
Variable & Type & Transformation\\
\midrule
Age range & Categorical & One-hot encode\\
Sex & Categorical & One-hot encode\\
Vehicle category & Categorical & One-hot encode\\
Make & Categorical & Embed in $\mathbb{R}^2$\\
Vehicle age & Numeric & Center and scale\\
Region & Categorical & Embed in $\mathbb{R}^2$\\
\bottomrule
\end{tabular}
\end{table}

\hypertarget{model-dev}{%
\section{Model Development}\label{model-dev}}

In this appendix, we describe the ML model and the data used to train it. Note
that, for our paper, the ultimate goal of the modeling procedure is to develop
something that can produce predictions. As a result, we do not follow standard
practices for tuning and validation. However, for the sake of
completeness and reproducibility, we include an overview of the process here.
Implementation is done using the R (R Core Team 2019) interface to TensorFlow
(Abadi et al. 2015). The model explanation visualizations utilize the
implemention by Biecek and Burzykowski (2019), and the code to reproduce them are available on
GitHub\footnote{https://github.com/kasaai/explain-ml-pricing}.

\hypertarget{data}{%
\subsection{Data}\label{data}}

We use data from the AUTOSEG (``Automobile Statistics System'') of Brazil's
Superintendence of Private Insurance (SUSEP). The organization maintains
policy-characteristics-level data, including claim counts and amounts, for all
insured automobiles in Brazil. The data contains variables from policyholder
characteristics to losses by peril. We use the records from the first half of
2012, which contains 1,707,651 records. One-fifth of the data is reserved for
testing; the remainder is further split into 3/4 of analysis and 1/4 into
assessment for determining early stopping.

\hypertarget{model}{%
\subsection{Model}\label{model}}

Table \ref{tab:variables} shows the input variables to our model and their
associated transformations. For ``make'' and ``region'', we map each level to a
point in \(\mathbb{R}^2\) through embedding layers. The model predicts expected
loss cost for all perils combined. The architecture is a feedforward
neural network with two hidden layers with 64 units each. The activations for
the hidden layers are ReLU while for the output layer it is softplus. Exposures
for each record are used as sample weights during training. We fit the model
via ADAM with an initial learning rate of 0.1, a mini-batch size of 10,000,
and trigger early stopping when the mean squared error on the asssessment set
does not improve for five epochs.

\end{document}